\documentclass[showpacs,twocolumn,aps,floatfix]{revtex4-1}%
\usepackage{amssymb}
\usepackage{amsfonts}
\usepackage{amsmath}
\usepackage{graphicx}%
\providecommand{\U}[1]{\protect\rule{.1in}{.1in}}

\begin{document}
\preprint{cond-mat}
\title[Short title for running header]{Large deviations in boundary-driven systems: Numerical evaluation and
effective large-scale behavior}
\author{Guy Bunin}
\affiliation{The Technion -- Israel Institute of Technology, Haifa 32000, Israel}
\author{Yariv Kafri}
\affiliation{The Technion -- Israel Institute of Technology, Haifa 32000, Israel}
\author{Daniel Podolsky}
\affiliation{The Technion -- Israel Institute of Technology, Haifa 32000, Israel}

\begin{abstract}
We study rare events in systems of diffusive fields driven out of equilibrium
by the boundaries. We present a numerical technique and use it to calculate
the probabilities of rare events in one and two dimensions. Using this
technique, we show that the probability density of a slowly varying
configuration can be captured with a small number of long wave-length modes.
For a configuration which varies rapidly in space this description can be
complemented by a local equilibrium assumption.

\end{abstract}
\volumeyear{year}
\volumenumber{number}
\issuenumber{number}
\eid{identifier}
\startpage{1}
\endpage{6}
\maketitle

In many cases the typical size of fluctuations in a physical system with $N$
degrees of freedom is of order $1/\sqrt{N}$. Larger fluctuations are rare, and
their probability scales as $P\left[  \rho\right]  \sim\exp\left(
-N\phi\left[  \rho\right]  \right)  \,$, where $\phi$ is an intensive function
of the state $\rho$. The function $\phi\left[  \rho\right]  $ is known as the
\emph{large-deviation} \emph{function} (LDF) and is of fundamental interest in
statistical mechanics. In equilibrium systems, $\phi$ is equal to the
free-energy density. Away from equilibrium, a simple expression for $\phi$ is
in general not known, and it may be affected by details of the system's
dynamics. Besides its fundamental interest for non-equilibrium physics, the
function $\phi$ is important in various applications, e.g. for calculating
escape rates from metastable states, with applications ranging from chemistry
and population dynamics to cosmology
\cite{transition_path,population_dynamics,colman,Noise_non_lin_book,analog_exp,Maier_Stein,Touchette}%
.

In a non-equilibrium steady-state, to compute the probability of a rare-event,
one must calculate the dynamics leading up to that event
\cite{Freidlin_Wentzell}. This is in general a difficult task, even more so
for spatially extended systems, where only a handful of analytical solutions
exist \cite{Derrida_review}. If a general understanding is to emerge,
additional methods beyond exact solutions need to be considered. Indeed,
recent years have seen a considerable effort to develop numerical techniques
to calculate the LDF in a variety of systems
\cite{Giardina,TV_simulation_cont_t,Giardina_review,current_large_dev_review,MAM_fin_dim}%
.

In this Letter, we study the LDF in bulk-conserving diffusive systems, which
are driven out of equilibrium by the boundaries. These describe a broad range
of transport phenomena, including electronic systems, ionic conductors, and
heat conduction \cite{shot_noise,KMP}. We show that the LDF in such systems
can be efficiently evaluated numerically for a \emph{general} interacting
system in one and two dimensions, giving us access to previously unavailable
information. This is done by searching for the most probable history
$\rho\left(  x,t\right)  $ of the conserved density function $\rho$ leading to
a rare state $\rho_{f}\left(  x\right)  $. Importantly, using the numerical
technique we show that for many non-trivial cases, the LDF of a slowly varying
configuration $\rho_{f}\left(  x\right)  $ can be calculated by considering
only histories $\rho\left(  x,t\right)  $ which are slowly varying in space,
i.e. which are given by the sum of only a few long wave-length modes. This
implies that the long wave-length structure of the LDF can be understood using
an \emph{effective finite-dimensional theory}, instead of the full infinite
dimensional one. In addition, we find that a local equilibrium assumption can
capture much of the short wave length structure. This could suggest a simple
framework to treat the LDF in these systems.

In bulk conserving diffusive systems, the conserved density $\rho\left(
\mathbf{x},t\right)  $, representing e.g. charge or energy density, is related
to the current $\mathbf{J}\left(  \mathbf{x},t\right)  $ by%
\begin{equation}
\partial_{t}\rho+\mathbf{\nabla}\cdot\mathbf{J}=0\ , \label{eq:conserve}%
\end{equation}
where the current is given by
\begin{equation}
\mathbf{J=-}D\left(  \rho\left(  \mathbf{x},t\right)  \right)  \mathbf{\nabla
}\rho\left(  \mathbf{x},t\right)  +\sqrt{\sigma\left(  \rho\left(
\mathbf{x},t\right)  \right)  }\boldsymbol{\eta}\left(  \mathbf{x},t\right)
\ . \label{eq:J_def}%
\end{equation}
$D\left(  \rho\left(  x,t\right)  \right)  $ is a density-dependent
diffusivity function, while $\sigma\left(  \rho\left(  \mathbf{x},t\right)
\right)  $ controls the size of the white noise $\boldsymbol{\eta}\left(
\mathbf{x},t\right)  $, which in $d$ dimensions satisfies $\left\langle
\eta_{a}\left(  \mathbf{x},t\right)  \right\rangle =0$ and $\left\langle
\eta_{a}\left(  \mathbf{x},t\right)  \eta_{b}\left(  \mathbf{x}^{\prime
},t^{\prime}\right)  \right\rangle =N^{-1}\delta_{ab}\delta^{d}\left(
\mathbf{x}-\mathbf{x}^{\prime}\right)  \delta\left(  t-t^{\prime}\right)  $.
The prefactor $N^{-1}$ in the noise variance results from the fact that we
have scaled distances by $L=N^{1/d}$ the system size, and time by $L^{2}$.
After this rescaling the noise is small as a consequence of the coarse
graining. $D\left(  \rho\right)  $ and $\sigma\left(  \rho\right)  $ are
related via a fluctuation-dissipation relation (Nyquist noise in electronic
systems), which for particle systems reads $\sigma\left(  \rho\right)
=2k_{B}T\rho^{2}\kappa\left(  \rho\right)  D\left(  \rho\right)  $ where
$\kappa\left(  \rho\right)  $ is the compressibility \cite{Derrida_review}.
Here we study a system on a domain $A$ connected to reservoirs which fix the
density at the boundary $\partial A$,
\begin{equation}
\left.  \rho\left(  \mathbf{x},t\right)  \right\vert _{\mathbf{x}\in\partial
A}=\left.  \rho_{B}\left(  \mathbf{x}\right)  \right\vert _{\mathbf{x}%
\in\partial A}\ . \label{eq:boundaries}%
\end{equation}
If the boundary density is not constant a current is induced through the
system, driving it out of equilibrium.

The average, or most probable density profile for the system $\bar{\rho}$, is
obtained by solving $\mathbf{\nabla}\cdot\left[  D\left(  \bar{\rho}\right)
\mathbf{\nabla}\bar{\rho}\right]  =0$, with $\bar{\rho}\left(  \mathbf{x}%
\right)  =\rho_{B}\left(  \mathbf{x}\right)  $ at the boundaries. In
equilibrium (i.e. when $\rho_{B}$ is constant), the steady-state probability
of any other density profile $\rho\left(  \mathbf{x}\right)  $ is easy to
obtain: the large deviation functional $\phi\left[  \rho\right]  $ is then
given by the free-energy, which is local in $\rho$. By contrast, the
steady-state probability distribution away from equilibrium is notoriously
hard to compute. In general, it is known that despite the local nature of the
dynamics, $\phi\left[  \rho\right]  $ is non-local in $\rho$, leading to
generic long-range correlations \cite{Derrida_review}. Analytical results for
$\phi\left[  \rho\right]  $ are known only for a few models of interacting
systems, corresponding to specific choices of $D\left(  \rho\right)  $ and
$\sigma\left(  \rho\right)  $, and almost exclusively in one dimension (1d).
The only known example in higher dimensions is the zero-range process, which
admits a trivial product measure \cite{ZRP,BertiniPRL}. Hence, despite the
central role that $\phi\left[  \rho\right]  $\ plays in the understanding of
non-equilibrium phenomena, little is known about its properties.

To compute the large deviation for the model described above, we first note
that the probability of a noise realization $\boldsymbol{\eta}\left(
\mathbf{x},t\right)  $ is Gaussian, $P\sim\exp\left(  -N\frac{1}{2}%
\int\boldsymbol{\eta}^{2}d\mathbf{x}dt\right)  $. Using this expression
together with Eq. (\ref{eq:J_def}), the probability of a history $\left\{
\rho\left(  \mathbf{x},t\right)  ,\mathbf{J}\left(  \mathbf{x},t\right)
\right\}  $\ during time $\tau\leq t\leq0$ is $P\sim\exp\left(  -NS\right)  $,
where the action $S$ is given by
\begin{equation}
S=\int_{\tau}^{0}dt\int_{A}d\mathbf{x}\frac{\left[  \mathbf{J}\left(
\mathbf{x},t\right)  +D\left(  \rho\left(  \mathbf{x},t\right)  \right)
\mathbf{\nabla}\rho\left(  \mathbf{x},t\right)  \right]  ^{2}}{2\sigma\left(
\rho\left(  \mathbf{x},t\right)  \right)  }\ .\label{eq:action}%
\end{equation}
As the noise is small, the system spends most of the time close to $\bar{\rho
}$, the unique fixed point of the diffusion equation at zero noise. To
calculate the steady-state probability of a rare event, $\rho_{f}\left(
\mathbf{x}\right)  $, we consider trajectories $\rho\left(  \mathbf{x}%
,t\right)  $ starting from $\bar{\rho}$ in the distant past $t\rightarrow
-\infty$ and ending at $\rho_{f}\left(  \mathbf{x}\right)  $ at $t=0$. For
large $N$, its probability $P\sim\exp\left\{  -N\phi\left[  \rho_{f}\right]
\right\}  $ is given by
\cite{BertiniPRL,BertiniJstat,JSP_quantum,TKL_long,Freidlin_Wentzell}%
\begin{equation}
\phi\left[  \rho_{f}\right]  =\inf_{\rho,\mathbf{J}}S\ ,\label{eq:phi_min_S}%
\end{equation}
where the infimum is over histories satisfying Eq. (\ref{eq:conserve}), with
initial and final conditions $\rho\left(  \mathbf{x},t\rightarrow
-\infty\right)  =\bar{\rho}\left(  \mathbf{x}\right)  $, $\rho\left(
\mathbf{x},t=0\right)  =\rho_{f}\left(  \mathbf{x}\right)  $, and the boundary
conditions, Eq. (\ref{eq:boundaries}). We now describe a numerical method
which utilizes this formulation to evaluate the large deviation $\phi\left[
\rho_{f}\right]  $.

To calculate numerically the large deviation, we directly minimize the action,
Eq. (\ref{eq:action}). We present a simple algorithm, which efficiently finds
minima of the action for such systems. The algorithm is based on starting with
a problem where the solution is known and gradually modifying it while
maintaining a minimizing solution. Specifically, we start with a problem where
the initial and final states are identical, $\rho\left(  \mathbf{x}%
,t\rightarrow\tau\right)  =\rho\left(  \mathbf{x},t=0\right)  =\bar{\rho
}\left(  \mathbf{x}\right)  $. In the exact solution $\tau=-\infty$; here we
take a sufficiently early time $\tau$, before any significant evolution has
begun, and check for convergence. Eq. (\ref{eq:action}) has a unique minimum
to this problem, $\rho\left(  \mathbf{x},t\right)  =\bar{\rho}\left(
\mathbf{x}\right)  $ and $\mathbf{J=-}D\left(  \bar{\rho}\right)
\mathbf{\nabla}\bar{\rho}$, for which the action vanishes, hence $\phi\left[
\bar{\rho}\right]  =0$. We now gradually change the final condition: define a
series of gradually changing profiles $\rho^{\left(  m\right)  }\left(
\mathbf{x}\right)  $, $m=1..n$, with $\rho^{\left(  1\right)  }=\bar{\rho}$,
and $\rho^{\left(  n\right)  }=\rho_{f}$. We call the series $\left\{
\rho^{\left(  m\right)  }\right\}  $ the \emph{final-state trajectory}. The
solution for the problem with $\rho^{\left(  1\right)  }\left(  \mathbf{x}%
\right)  $ is known. In the next iteration we solve the same problem as above,
only with final conditions $\rho\left(  \mathbf{x},t=0\right)  =\rho^{\left(
2\right)  }\left(  \mathbf{x}\right)  $, using as an initial guess
$\rho\left(  \mathbf{x},t\right)  $ as obtained from the previous iteration.
This procedure is iterated until we reach the final condition $\rho^{\left(
n\right)  }\left(  \mathbf{x}\right)  =\rho_{f}\left(  \mathbf{x}\right)  $.
Standard algorithms can be used for the minimization at each step; we have
experimented with different algorithms, and the final results did not depend
on this choice, but computational efficiency did. Since $S$ is a sum of
squares, non-linear least-squares methods are applicable and were found to be efficient.

Within each iteration, we minimize the action $\inf_{\rho,\mathbf{J}}S\left[
\rho,\mathbf{J}\right]  =\inf_{\rho}\tilde{S}\left[  \rho\right]  $, where
$\tilde{S}\left[  \rho\right]  \equiv\inf_{\mathbf{J}}S\left[  \rho
,\mathbf{J}\right]  $. To evaluate $\tilde{S}\left[  \rho\right]  $ we take
into account the constraint Eq. (\ref{eq:conserve}), by introducing a Lagrange
multiplier $\hat{\rho}\left(  \mathbf{x},t\right)  $, and optimizing
$G=S+\int\hat{\rho}\left(  \mathbf{x},t\right)  \left(  \partial_{t}%
\rho+\mathbf{\nabla}\cdot\mathbf{J}\right)  d\mathbf{x}dt$ with respect to
$\mathbf{J}$ and $\hat{\rho}$. $\delta G/\delta\mathbf{J}=0$ gives
$\mathbf{J}=$ $-D\left(  \rho\right)  \mathbf{\nabla}\rho+\sigma\left(
\rho\right)  \mathbf{\nabla}\hat{\rho}$, which together with Eq.
(\ref{eq:conserve}) reads
\begin{equation}
\partial_{t}\rho=\mathbf{\nabla}\cdot\left(  D\left(  \rho\right)
\mathbf{\nabla}\rho\right)  -\mathbf{\nabla}\cdot\left(  \sigma\left(
\rho\right)  \mathbf{\nabla}\hat{\rho}\right)  \ , \label{eq.Hamilton_rho_EOM}%
\end{equation}
with boundary conditions $\left.  \hat{\rho}\right\vert _{\mathbf{x}%
\in\partial A}=0$ \cite{BertiniJstat}. This is a linear equation for
$\hat{\rho}\left(  \mathbf{x},t\right)  $ in terms of $\rho\left(
\mathbf{x},t\right)  $, which can easily be solved numerically. \ Substituting
the expression for $\mathbf{J}$ into Eq. (\ref{eq:action}), we find that
$\tilde{S}\left[  \rho\right]  =\frac{1}{2}\int dtd\mathbf{x}\sigma\left(
\rho\right)  \left(  \mathbf{\nabla}\hat{\rho}\right)  ^{2}$. In practice this
is carried out numerically by discretizing the equations, as is described in
detail in the Appendix. For reference below, we note that on the minimal path
$\delta G\left(  \rho,J\right)  /\delta\rho=0$ also holds, which yields an
equation of motion for $\hat{\rho}$:%
\begin{equation}
\partial_{t}\hat{\rho}=-D\left(  \rho\right)  \nabla^{2}\hat{\rho}-\frac{1}%
{2}\sigma^{\prime}\left(  \rho\right)  \left(  \mathbf{\nabla}\hat{\rho
}\right)  ^{2} \label{eq.Hamilton_rho_hat_EOM}%
\end{equation}
Note that the $\hat{\rho}$~field is the momentum conjugate to $\rho$ in a
Hamiltonian formulation of the problem \cite{BertiniJstat}.

In the context of boundary driven-diffusive systems numerical techniques were
used to study current large deviations (and similar quantities)
\cite{Giardina,TV_simulation_cont_t,Giardina_review,current_large_dev_review,hurtado_PNAS,hurtado2}%
. Algorithms have also been devised to calculate generating functions in
related systems \cite{Kamenaev}. However, both of these quantities do not
yield direct information on the probability density at a specific
state.\textbf{\ }Our algorithm directly minimizes the action, as do
\cite{MAM_fin_dim}. A key feature of our algorithm is the gradual change of
the final state, which makes it both stable and insensitive to the choice of
optimization algorithm. The algorithm is easy to implement.\ As a further
advantage, the algorithm is easily modified to use only a small number of
modes, as explored below.

The numerics were tested against the known 1d and two dimensional (2d) models
for which analytical expressions for large deviation and the trajectory
minimizing the action exist. As a first demonstration, we consider the simple
symmetric exclusion process (SSEP) in 1d
\cite{BertiniJstat,SSEP_spohn,SSEP_large_dev}.\ The model describes a lattice
gas with hard-core exclusion, and in the continuum limit leads to $D=1$,
$\sigma\left(  \rho\right)  =2\rho\left(  1-\rho\right)  $ with $0\leq\rho
\leq1$. We take the domain $0<x<1$, and the boundary conditions are $\rho
_{B}\left(  0\right)  ,\rho_{B}\left(  1\right)  $. When $\rho_{B}\left(
0\right)  \neq\rho_{B}\left(  1\right)  $ the system is driven out of
equilibrium. The most probable state is $\bar{\rho}\left(  x\right)  =\left(
1-x\right)  \rho_{B}\left(  0\right)  +x\rho_{B}\left(  1\right)  $.

\begin{figure}
[ptb]
\begin{center}
\includegraphics[
trim=0.000000in 0.000000in 0.004360in 0.000000in,
height=1.7181in,
width=2.5075in
]%
{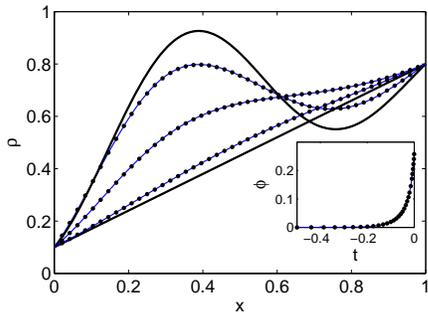}%
\caption{Optimal trajectory in the 1d SSEP for a given final state $\rho_{f}$.
Solid lines are $\bar{\rho},\rho_{f}$ and the exact $\rho\left(  x,t\right)  $
at $t=-0.01,-0.05,-0.2$. Numerical results are shown as dots. Inset: the
contribution to the action up to intermediate times.}%
\label{fig:ssep_numerics}%
\end{center}
\end{figure}

Fig. \ref{fig:ssep_numerics} shows an example of a path $\rho\left(
x,t\right)  $ minimizing the action for the SSEP, with a given final state
$\rho_{f}$. In this case it is known \cite{SSEP_large_dev} that there is a
unique local minimizer for the action, and indeed we find a single solution,
independent of the different final state trajectories tested.\ Shown is the
trajectory from the initial state $\bar{\rho}$ to the final state $\rho_{f}$
at different times, compared with the numerical solution, showing close
agreement. The inset shows the contribution to the large deviation (the action
integral, Eq. (\ref{eq:action})), integrated up to time $t$. The numerics
where carried out with $N_{x}=48$ space divisions and $N_{t}=30$ time
divisions. In order to capture the time evolution more exactly, the size of
the time intervals $\Delta t_{i}=t_{i+1}-t_{i}$ was taken to be a geometric
series, where the last division is $40$ times smaller than the first. The
initial time was $\tau=-0.5$. As is clear from the inset, contributions from
earlier time are negligible. The relative error in the large deviation is
$4\cdot10^{-3}$. By taking $N_{x}=150,N_{t}=70$ the error is reduced to
$10^{-3}$.
\ 

We now show that the dynamics of the large deviations of slowly varying
configurations can be \emph{captured by a small number of variables}, which
describe the long wave-length behavior. This provides an effective large-scale
description of $\phi$ in terms of a small number of degrees of freedom.\ To
this end we define a family of functions $\left\{  \rho_{i}\left(
\mathbf{x}\right)  \right\}  $ which span the function space, ordered such
that $\rho_{1}\left(  \mathbf{x}\right)  $ is the slowest varying in space,
followed by $\rho_{2}\left(  \mathbf{x}\right)  $, etc.. We then consider an
approximation in which the configurations leading to $\rho_{f}$ are restricted
to be linear combinations of a finite number of the slowest-varying $\rho_{i}%
$,%
\begin{equation}
\rho_{M}\left(  x,t\right)  =\bar{\rho}\left(  x\right)  +\sum_{i=1}^{M}%
a_{i}\left(  t\right)  \rho_{i}\left(  x\right)  \ .\label{eq.finite_sum}%
\end{equation}
For $M\rightarrow\infty$ this recovers the exact extremal solutions. For
finite $M$, when $\rho_{f}$ is itself of the form of Eq. (\ref{eq.finite_sum})
the solutions give upper bounds for the exact value of $\phi\left[  \rho
_{f}\right]  $, since the minimization is only on a subset all histories
$\rho\left(  x,t\right)  $. Below we are interested in how well they
approximate the exact solutions.

\begin{figure}
[ptb]
\begin{center}
\includegraphics[
height=1.6561in,
width=2.5581in
]%
{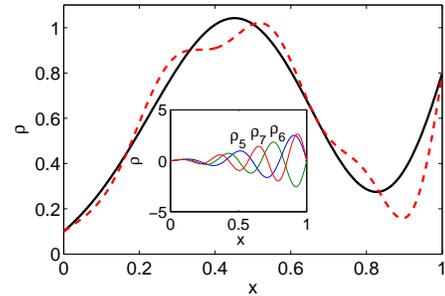}%
\caption{A profile containing only the two lowest KMP\ modes (solid), and one
contaning modes up to 7 modes (dashed). Inset: plot of KMP modes 5,6 and 7.}%
\label{fig:local_eq}%
\end{center}
\end{figure}

A natural choice of the functions $\left\{  \rho_{i}\left(  x\right)
\right\}  $ are the normal modes of the linearized Hamilton evolution, Eqs.
(\ref{eq.Hamilton_rho_EOM}) and (\ref{eq.Hamilton_rho_hat_EOM}), linearized
around $\rho=\bar{\rho}$ and $\hat{\rho}=0$ \cite{other_modes}. These are
obtained by substituting $\rho\left(  x,t\right)  =\bar{\rho}\left(  x\right)
+\sum_{i}\rho_{i}\left(  x\right)  e^{\lambda_{i}t}$, $\hat{\rho}\left(
x,t\right)  =\sum_{i}\hat{\rho}_{i}\left(  x\right)  e^{\lambda_{i}t}$ and
keeping only linear terms in $\rho_{i}\left(  x\right)  $ and $\hat{\rho}%
_{i}\left(  x\right)  $
\begin{subequations}
\begin{align}
\lambda_{i}\rho_{i} &  =\mathbf{\nabla}\cdot\left(  D\left(  \bar{\rho
}\right)  \mathbf{\nabla}\rho_{i}\right)  -\mathbf{\nabla}\cdot\left(
\sigma\left(  \bar{\rho}\right)  \mathbf{\nabla}\hat{\rho}_{i}\right)
\ ,\label{eq:modes_EOM1}\\
\lambda_{i}\hat{\rho}_{i} &  =-D\left(  \bar{\rho}\right)  \nabla^{2}\hat
{\rho}_{i}\ ,\label{eq:modes_EOM2}%
\end{align}
with boundary conditions $\rho_{i}=0=\hat{\rho}_{i}$. The solution
$\rho\left(  x,t\right)  $ of these equations is the optimal trajectory of Eq.
(\ref{eq:action}) for small fluctuations of $\rho$ around $\bar{\rho}$. These
equations admit two types of solutions. In one type, $\hat{\rho}_{i}$ vanish
identically. These solutions correspond to the zero-noise evolution, and do
not satisfy the initial condition at $t\rightarrow-\infty$ (except in the
trivial case $\rho_{f}=\bar{\rho}$). The other set of solutions is obtained by
first solving Eq. (\ref{eq:modes_EOM2}), which is an eigenvalue problem for
$\hat{\rho}_{i}$, independently of $\rho_{i}$. The resulting $\lambda_{i}%
,\hat{\rho}_{i}$ are then substituted into Eq. (\ref{eq:modes_EOM1}), and
$\rho_{i}\left(  x\right)  $ is solved for. As a convention, we take all
$\rho_{i}$ functions to be normalized with $\int\rho_{i}^{2}dx=1$, and in 1d
have a positive slope at $x=0$.

As a first example we return to the profile $\rho_{f}$ given in Fig.
\ref{fig:ssep_numerics}, which is of the form $\rho_{f}=\bar{\rho}+\alpha
_{1}\rho_{1}+\alpha_{2}\rho_{2}$, where $\lambda_{1,2}$ are the two lowest
eigenvalues, $\lambda_{i}=\left(  \pi i\right)  ^{2}$, and $\alpha_{1}%
=-0.256$,\ $\alpha_{2}=-0.214$. The modes $\rho_{i}$, for $i=1,2,3$ are shown
in Fig. \ref{ssep_1d_modes_err_evolution_v5}(a). We now minimize the action
with histories constrained to be of the form of Eq. (\ref{eq.finite_sum}),
with $M=2,3,..$. Fig. \ref{ssep_1d_modes_err_evolution_v5}(b) shows the
histories for $M=2$ (dashed lines). Even for $M=2$, the large deviation is
obtained to within 2\%, see Fig. \ref{ssep_1d_modes_err_evolution_v5}(c),
suggesting that even at this level the system's behavior can be captured by an
effective model with only \emph{two degrees of freedom}. Such a small error is
striking considering the highly non-linear nature of the problem (as $\rho
_{f}$ is far from $\bar{\rho}$), which is generically expected to mix higher
modes in significant amounts. By comparison, a local equilibrium approximation
(where a space-dependent chemical potential is set to reproduce $\bar{\rho}$)
gives a relative error of 16\%, whereas extending the linearized dynamics to
the full evolution \cite{SSEP_spohn} gives an error in the large deviation of
67\% (mostly due to the long-range Gaussian corrections).\ This highlights
both the importance of non-linearities in this problem, and the success of the
truncated approximation.

The mode approximation can also be used as a high precision numerical method.
As shown in Fig. \ref{ssep_1d_modes_err_evolution_v5}(c), for $M=15$, the
relative error is reduced to $\sim10^{-5}$, well below the error obtained by a
straightforward discretization of space. Indeed similar approaches have been
used as numerical tools to improve accuracy in \cite{MAM_fin_dim}.
\begin{figure}
[ptb]
\begin{center}
\includegraphics[
trim=0.000000in 0.000000in 0.000000in -0.215831in,
height=2.2308in,
width=3.1284in
]%
{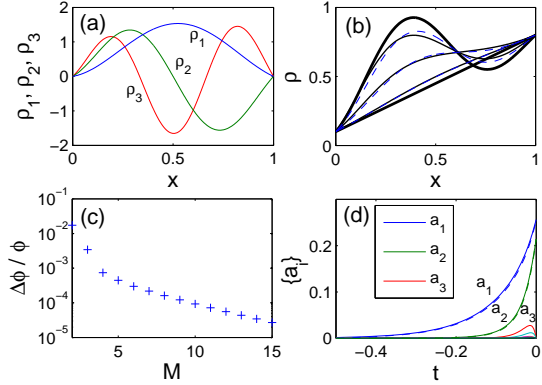}%
\caption{(a) The modes $\rho_{1},\rho_{2},\rho_{3}$. (b) The exact evolution
(solid lines) and the evolution with $M=2$ (dashed lines). (c) The relative
error $\left(  \phi_{M}\left[  \rho_{f}\right]  -\phi\left[  \rho_{f}\right]
\right)  /\phi\left[  \rho_{f}\right]  $ for different numbers of modes $M$.
(d) The time evolution of $a_{i}$ for $i=1..9$. Modes $a_{5}$ and above cannot
be seen on this scale.}%
\label{ssep_1d_modes_err_evolution_v5}%
\end{center}
\end{figure}

The low mode approximation is also useful in cases where an exact analytical
solution does not exist. For example, our method allows us to study
two-dimensional (2d) systems. In 2d analytic solutions are only known for a
set of models which exhibit no long-range correlations. We used one such
model, with $D=1$ and $\sigma\left(  \rho\right)  =2\rho$, corresponding to a
model of non-interacting particles, as a benchmark for our method, and found
agreement between the numerics and the exact solution (see the Appendix). In
what follows we present results for an interacting system, the SSEP in 2d, for
which an exact expression for the large deviation is not known. This exhibits
the real power of the numerics. To show the generality of the low mode
approximation, we take a somewhat arbitrary choice of boundary conditions,
$\rho\left(  x,y\right)  =0.25\sin(3\operatorname{atan}\left(  y/x\right)
)+0.55+0.5y$, on the square domain $A=\left[  -1/2,1/2\right]  ^{2}$, see Fig.
\ref{fig:2d_ssep_combined}. The most probable density profile $\bar{\rho}$\ is
shown in Fig. \ref{fig:2d_ssep_combined}(a), and we present results for the
profile $\rho_{f}$ shown in Fig. \ref{fig:2d_ssep_combined}(b), which, as in
the 1d discussion, is of the form $\rho_{f}=\bar{\rho}+\alpha_{1}\rho
_{1}+\alpha_{2}\rho_{2}$, where $\rho_{1},\rho_{2}$ are the lowest modes in
this system, and $\alpha_{1}=0.097,\alpha_{2}=0.128$. Similar results were
obtained for other profiles.\ Fig. \ref{fig:2d_ssep_combined}(c) shows the
growth of the modes for $M=2..10$. The first two modes give the exact large
deviation to within $2\cdot10^{-3}$, as estimated by $\left(  \phi_{2}\left[
\rho_{f}\right]  -\phi_{14}\left[  \rho_{f}\right]  \right)  /\phi_{14}\left[
\rho_{f}\right]  $, see Fig. \ref{fig:2d_ssep_combined}(d). Once more, as in
1d, this means that the evolution is well described by a two-parameter space
$\left(  a_{1}\left(  t\right)  ,a_{2}\left(  t\right)  \right)  $, despite
the non-linear nature of the problem. Interestingly, a local equilibrium
approximation gives a relatively low error of\ 1.2\%.%
\begin{figure}
[ptb]
\begin{center}
\includegraphics[
height=2.4161in,
width=3.2552in
]%
{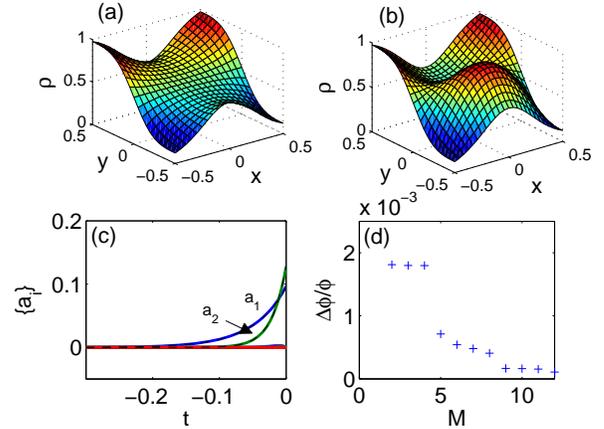}%
\caption{(a) $\bar{\rho}$ and (b) $\rho_{f}$ for a 2d SSEP example. (c) The
modes evolution. Solid lines: Amplitudes $a_{i}$ for $i=1..10$. $a_{3}$ and
above are hard to distinguish in the scale of the graph.\ Dashed lines (almost
indistinguishable): 2 modes. (d) The relative error $\left(  \phi_{M}\left[
\rho_{f}\right]  -\phi\left[  \rho_{f}\right]  \right)  /\phi\left[  \rho
_{f}\right]  $.}%
\label{fig:2d_ssep_combined}%
\end{center}
\end{figure}

We have shown so far that the large deviation function is well-approximated
using only a few modes, provided that $\rho_{f}$ itself is a slowly-varying
function of space. Here we discuss how to extend these results to cases where
$\rho_{f}$\ is not necessarily slowly varying, but has some high-mode content.
Recalling that the bulk behavior is governed by equilibrium dynamics, one
might expect that small, high-mode perturbations around the low mode behavior
would be captured by a local equilibrium theory. In particular, define a local
free-energy density $f_{eq,\bar{\rho}}\left(  \rho\right)  \equiv\int
_{\bar{\rho}}^{\rho}d\rho^{\prime}\int_{\bar{\rho}}^{\rho^{\prime}}%
\frac{2D\left(  \rho^{\prime\prime}\right)  }{\sigma\left(  \rho^{\prime
\prime}\right)  }d\rho^{\prime\prime}$. In equilibrium, when all boundary
densities are equal, this is precisely the free-energy density. We then
expect
\end{subequations}
\begin{equation}
\phi\left[  \rho_{f}\right]  -\phi\left[  \rho_{f,M}\right]  \simeq\int\left[
f_{eq,\bar{\rho}}\left(  \rho_{f}\right)  -f_{eq,\bar{\rho}}\left(  \rho
_{f,M}\right)  \right]  dx\ ,\label{eq:local_eq}%
\end{equation}
where $\rho_{f,M}$ is $\rho_{f}$ projected to the subspace spanned by
$\left\{  \bar{\rho},\rho_{1},..,\rho_{M}\right\}  $. In other words, the
error due to truncation of the high modes is approximately accounted for by a
local equilibrium theory. We now show that this is indeed the case in an
example on the Kipnis--Marchioro--Presutti (KMP) model for heat transfer
\cite{KMP}, whose continuum limit \cite{KMP_large_dev} gives $D=1$ and
$\sigma\left(  \rho\right)  =2\rho^{2}$ (similar results are obtained for
other models). In Fig. \ref{fig:local_eq}, we take $\rho_{f}$ of the form
$\bar{\rho}+0.463\rho_{1}+0.507\rho_{2}-0.1\left(  \rho_{5}+\rho_{6}+\rho
_{7}\right)  $. For this profile, the LHS of Eq. (\ref{eq:local_eq})\ for
$M=2$ equals -0.03667 and the RHS equals -0.03623. Thus Eq. (\ref{eq:local_eq}%
)\ is satisfied with relative accuracy of 1.2\%. Hence, for profiles with high
mode content the effective low mode description can be corrected for using Eq.
(\ref{eq:local_eq}).%

In summary, our findings suggest that the evolution leading to a
\textquotedblleft smooth\textquotedblright\ rare event is smooth: the
continuous diffusion $\partial_{t}\rho=\mathbf{\nabla}\cdot\left(  D\left(
\rho\right)  \mathbf{\nabla}\rho\right)  +\left\langle
small\ noise\right\rangle $, makes an enduring high frequency perturbation
highly unlikely. Indeed we show that even when the evolution leading to the
rare event is restricted to profiles involving just a few modes, good
quantitative agreement with the exact large deviation may be found. This means
that the large deviation is well-described in a space involving just a few variables.

Finally, we note that in general the action may have more than a single local
minimum, an effect which is well-known in finite-dimensional systems
\cite{Noise_non_lin_book,analog_exp,Maier_Stein,Graham_Tel,Dykman}. While this
does not happen in the well-known models used above in our comparisons, it
does in fact exist in other models, and can be studied using the numerical
method described here by considering various different final-state
trajectories. For example, when two solutions are present, different choices
of final-state trajectories will lead to the different local minima
\cite{local_min_footnote}. A detailed account of this issue and its physical
implications is beyond the scope of the present work, and will be disscussed
elsewhere \cite{ours}.

\section{Appendix\label{sec:appendix_A}}

\subsection{Action evaluation}

The action evaluation, including the Lagrange multipliers, is implemented
directly in the discrete setting, which gives discrete variants of Eq.
(\ref{eq.Hamilton_rho_EOM}), along with the boundary conditions. In the
numerical implementation, time and space are discretized, and $\rho$ is kept
at points $\rho_{i,k}=\rho\left(  x_{i},t_{k}\right)  $ in 1d, and
$\rho_{i,j,k}=\rho\left(  x_{i},y_{j},t_{k}\right)  $ in 2d. We start by
describing the method in 2d, and then discuss the simplifications which occur
in 1d.

The action, Eq. (\ref{eq:action}), is discretized as $S=\sum_{k}\Delta S_{k}$
where $\Delta S_{k}$ is the value of $S$ associated with the time interval
$\left[  t_{k},t_{k+1}\right]  $. This allows for the time resolution to vary.
For each $k$ separately, $\Delta S_{k}$ is evaluated as
\begin{equation}
\Delta S_{k}=\Delta t\sum_{i,j}\left(  S_{i+1/2,j}^{x}+S_{i,j+1/2}^{y}\right)
\ , \label{eq:Appendix_del_S}%
\end{equation}
where $\Delta t=t_{k+1}-t_{k}$. $S_{i+1/2,j}^{x}$ corresponds to the bond
connecting $\left(  i,j\right)  $ to $\left(  i+1,j\right)  $ (and similarly
for other half-integer indices).

Let $\rho_{i,j}\equiv\frac{1}{2}\left(  \rho_{i,j,k+1}+\rho_{i,j,k}\right)  $.
Then $S_{i+1/2,j}^{x}$ is given by%
\begin{equation}
S_{i+1/2,j}^{x}=\frac{\left(  J_{i+1/2,j}^{x}+D\left(  \rho_{i+1/2,j}\right)
\left(  \partial_{x}\rho\right)  _{i+1/2,j}\right)  ^{2}}{2\sigma\left(
\rho_{i+1/2,j}\right)  } \label{eq:Appendix_Sx}%
\end{equation}
with $\rho_{i+1/2,j}\equiv\frac{1}{2}\left(  \rho_{i,j}+\rho_{i+1,j}\right)
$, $\left(  \partial_{x}\rho\right)  _{i+1/2,j}\equiv\left(  \rho_{i+1,j}%
-\rho_{i,j}\right)  /\Delta x$, and a similar expression for $S_{i,j+1/2}^{y}%
$. The currents $J^{x},J^{y}$ are constrained to satisfy a discretized version
of the continuity equation, Eq. (\ref{eq:conserve}),
\begin{align}
0  &  =R_{ij}\equiv\left(  \partial_{t}\rho\right)  _{i,j}\nonumber\\
&  +\frac{J_{i+1/2,j}^{x}-J_{i-1/2,j}^{x}}{\Delta x}+\frac{J_{i,j+1/2}%
^{y}-J_{i,j-1/2}^{y}}{\Delta y}\ , \label{eq:Kirchoff}%
\end{align}
where $\left(  \partial_{t}\rho\right)  _{i,j}\equiv\left(  \rho
_{i,j,k+1}-\rho_{i,j,k}\right)  /\Delta t$, and $\Delta x,\Delta y$ are the
(constant) spacings in the $x$- and $y$-directions. To minimize the currents
subject to Eq. (\ref{eq:Kirchoff}), we define $\Delta G_{k}=\Delta S_{k}%
-\sum_{ij}\hat{\rho}_{ij}R_{ij}$. Differentiating $\Delta G_{k}$ with respect
to the currents gives%
\begin{align}
J_{i+1/2,j}^{x}  &  =-D\left(  \rho_{i+1/2,j}\right)  \left(  \partial_{x}%
\rho\right)  _{i+1/2,j}\label{eq:Appendix_J}\\
&  +\sigma\left(  \rho_{i+1/2,j}\right)  \frac{\hat{\rho}_{i+1,j}-\hat{\rho
}_{i,j}}{\Delta x}\nonumber
\end{align}
with a similar expression for $J_{i,j+1/2}^{y}$. This is a discrete variant of
$\mathbf{J}=$ $-D\left(  \rho\right)  \mathbf{\nabla}\rho+\sigma\left(
\rho\right)  \mathbf{\nabla}\hat{\rho}$. Substituting Eq. (\ref{eq:Appendix_J}%
) into Eq. (\ref{eq:Kirchoff}), one obtains a linear set of equations for the
$\hat{\rho}$-variables, which corresponds to Eq. (\ref{eq.Hamilton_rho_EOM}).
These are solved to find the $\hat{\rho}$-variables. Note that on boundary
sites Eq. (\ref{eq:Kirchoff}) involves only three currents (or two at corners
of the lattice), which is equivalent to setting $\hat{\rho}_{i,j}=0$ for $i,j$
outside the lattice. This corresponds to the boundary conditions $\left.
\hat{\rho}\left(  x,t\right)  \right\vert _{x\in\partial A}=0$ in the continuum.

Given the\ $\hat{\rho}$ values, the final expression for $S$ is obtained by
combining Eqs. (\ref{eq:Appendix_del_S}),(\ref{eq:Appendix_Sx}) and
(\ref{eq:Appendix_J}), and reads%
\begin{equation}
\Delta S_{k}=\frac{1}{2}\Delta t\Delta x\Delta y\sum_{i,j}\left[
\begin{array}
[c]{c}%
\sigma\left(  \rho_{i+1/2,j}\right)  \left(  \frac{\hat{\rho}_{i+1,j}%
-\hat{\rho}_{i,j}}{\Delta x}\right)  ^{2}\\
+\sigma\left(  \rho_{i,j+1/2}\right)  \left(  \frac{\hat{\rho}_{i,j+1}%
-\hat{\rho}_{i,j}}{\Delta y}\right)  ^{2}%
\end{array}
\right]  \ ,
\end{equation}
which serves as the discrete analog of $\tilde{S}\left[  \rho\right]
=\frac{1}{2}\int dtd\mathbf{x}\sigma\left(  \rho\right)  \left(
\mathbf{\nabla}\hat{\rho}\right)  ^{2}$. This concludes the evaluation of the
action $S$ for a given $\rho$. This procedure is used as a building block in
the optimization algorithm, where $S$ is evaluated for different histories
$\rho\left(  x,t\right)  $, see the main text.

In 1d the above scheme is somewhat simplified. Of course, only terms in the
$x$ direction appear. The continuity Eq. (\ref{eq:Kirchoff}) is now
$J_{i+1/2}^{x}=J_{i-1/2}^{x}-\Delta x\left(  \partial_{t}\rho\right)  _{i}$,
so $J_{i+1/2}^{x}=J_{c}-\Delta x\sum_{n=1}^{i}\,\left(  \partial_{t}%
\rho\right)  _{n}$, where $J_{c}$\ is independent of the position $i$ (but may
depend on time). Summing over Eq. (\ref{eq:Appendix_J}), and using $\hat{\rho
}_{0}=0$ we find
\begin{equation}
\hat{\rho}_{i}=\Delta x\sum_{m=0}^{i-1}\frac{D\left(  \rho_{m+1/2}\right)
\left(  \partial_{x}\rho\right)  _{m+1/2}+J_{c}-\Delta x\sum_{n=1}%
^{m}\,\left(  \partial_{t}\rho\right)  _{n}}{\sigma\left(  \rho_{m+1/2}%
\right)  }\ ,
\end{equation}
where $J_{c}$ is fixed by requiring that the boundary condition $\hat{\rho
}_{N_{x}+1}=0$ holds.

As an additional tool to improve accuracy, it is possible to interpolate
$\rho\left(  x,t\right)  $ onto a finer grid in $\left(  x,t\right)  $ before
evaluating the action. This simple step improves accuracy and stability at low
resolutions. In the example presented below, we use this technique to double
the time resolution.

Computations involving modes use exactly the same action evaluation scheme,
and only differ in the profiles $\rho\left(  x,t\right)  $ allowed in the
density optimization process.%
\begin{figure}
[h]
\begin{center}
\includegraphics[
trim=0.000000in 0.000000in 0.003809in 0.000000in,
height=1.6909in,
width=2.8078in
]%
{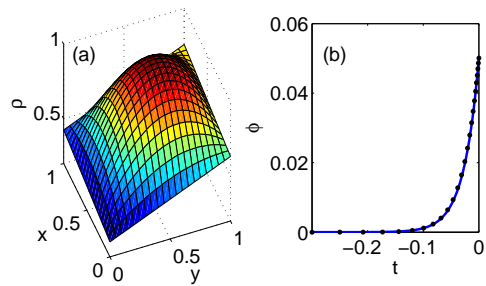}%
\caption{Comparison of numerical results with a 2d non-interacting model. (a)
The profile tested. (b) The exact evolution of the large deviation (solid
line) compared with the numerics (points).}%
\label{fig:ni_2d_low_rez}%
\end{center}
\end{figure}

\subsection{Benchmark for 2d numerical method}

The algorithm was tested in 2d against the model $\sigma=2\rho$ and
$D=1$.\ This model is a particular case of the open boundary Zero Range
process \cite{ZRP,BertiniPRL}, and its large deviation is given by%
\begin{equation}
\phi\left[  \rho_{f}\right]  =\int d^{2}x\left(  \rho_{f}\ln\frac{\rho_{f}%
}{\bar{\rho}}+\bar{\rho}-\rho_{f}\right)  \ .
\end{equation}
$\phi\left[  \rho_{f}\right]  $ was calculated for $\rho_{f}$ in Fig.
\ref{fig:ni_2d_low_rez}(a). Fig. \ref{fig:ni_2d_low_rez}(b) shows a comparison
of the numerical method with the exact result at a relatively low resolution,
with $N_{x}=N_{y}=8$ divisions in each space dimension and $N_{t}=25$
divisions in time, starting from $\tau=-0.3$. The profiles were interpolated
onto a grid with twice the time resolution before the action evaluation. The
relative error in $\phi\left[  \rho_{f}\right]  $ was $5\cdot10^{-4}$.


\begin{thebibliography}{99}                                                                                               %


\bibitem {transition_path}P. G. Bolhuis, D. Chandler, C. Dellago, P. L.
Geissler,\ Annu. Rev. Phys. Chem. 53, 291-318 (2002)

\bibitem {population_dynamics}A. Kamenev and B. Meerson, Phys. Rev. E. 77,
061107 (2008)

\bibitem {colman}S. Coleman, Phys. Rev. D 15, 2929--2936 (1977)

\bibitem {Noise_non_lin_book}F. Moss and P. V. E. McClintock, ed., Noise in
Nonlinear Dynamical Systems, Cambridge University Press, Cambridge (1989)

\bibitem {analog_exp}D. G. Luchinsky, P. V. E. McClintock, and M. I. Dykman,
Rep. Prog. Phys., 61(8):889-997 (1998)

\bibitem {Maier_Stein}R. S. Maier and D. L. Stein, Phys. Rev. E, 48(2):931-938 (1993)

\bibitem {Graham_Tel}R. Graham and T. T\'{e}l, Phys. Rev. Lett. 52, 9--12 (1984)

\bibitem {Dykman}M.I. Dykman, M.M. Millonas and V.N. Smelyanskiy, Phys. Lett.
A, 195 (1994), 53

\bibitem {Touchette}H. Touchette, R. J. Harris, Large deviation approach to
nonequilibrium systems, in R. Klages, W. Just, C. Jarzynski (eds),
Nonequilibrium Statistical Physics of Small Systems: Fluctuation Relations and
Beyond, Wiley-VCH, Weinheim (2012)

\bibitem {Freidlin_Wentzell}M. I.\ Freidlinand A. D. Wentzell, Random
Perturbations of Dynamical Systems, Springer-Verlag (1984)

\bibitem {Derrida_review}B. Derrida, J. Stat. Mech. P07023 (2007)

\bibitem {Giardina}C. Giardin\`{a}, J. Kurchan, and L. Peliti, Phys. Rev.
Lett. 96, 120603 (2006)

\bibitem {TV_simulation_cont_t}V. Lecomte and J. Tailleur, J. Stat. Mech.
P03004 (2007)

\bibitem {current_large_dev_review}J. Tailleur and V. Lecomte, AIP Conf. Proc.
1091, 212-219 (2008)

\bibitem {Giardina_review}C. Giardina, J. Kurchan, V. Lecomte, J. Tailleur, J.
Stat. Mech. 45 4 (2011)

\bibitem {MAM_fin_dim}D. Passerone and M. Parrinello, Phys. Rev. Lett. 87,
108302 (2001). W. E, W. Ren and E. Vanden-Eijnden, Commun. Pure Appl. Math.
57, 637 (2004). X. Wan, J. Comp. Phys. 230 (2011) 8669--8682. F. Bouchet, J.
Laurie and O. Zaboronski, J. Phys. Conf. Ser. 318 (2011) 022041

\bibitem {shot_noise}Ya. M. Blanter and M. B\"{u}ttiker, Physics Reports, 336,
1-2 (2000), W. Dieterich, P. Fulde and I. Peschel, Adv. in Phys. 29, 527-605 (1980)

\bibitem {KMP}C. Kipnis, C. Marchioro and E. Presutti, J. Stat. Phys. 27 65 (1982)

\bibitem {ZRP}C. Kipnis and C. Landim, Scaling Limits of Interacting Particle
Systems (Springer, New York, 1999)

\bibitem {BertiniPRL}L. Bertini, A. De Sole, D. Gabrielli, G. Jona-Lasinio,
and C. Landim, Phys. Rev. Lett. 87, 040601 (2001)

\bibitem {BertiniJstat}L. Bertini, A. De Sole, D. Gabrielli, G. Jona-Lasinio,
and C. Landim, J. Stat. Phys. 107, P07014 (2002)

\bibitem {JSP_quantum}A. N. Jordan, E. V. Sukhorukov and S. Pilgram, J. Math.
Phys. 45 4386-4417 (2004)

\bibitem {TKL_long}J. Tailleur, J. Kurchan and V. Lecomte, J. Phys. A: Math.
Theor. 41 505001 (2008)

\bibitem {hurtado2}P. I. Hurtado and P. L. Garrido, Phys. Rev. Lett. 107,
180601 (2011)

\bibitem {hurtado_PNAS}P. I. Hurtado et. al., Proc. Natl. Acad. Sci. USA.
108(19) 7704--7709 (2011)

\bibitem {Kamenaev}V. Elgart and A. Kamenev, Phys. Rev. E 70, 041106 (2004)

\bibitem {ours}G. Bunin, Y. Kafri, D. Podolsky, in preparation.

\bibitem {local_min_footnote}Every history which locally minimizes the action
will be found by a final state trajectory which follows this history.

\bibitem {SSEP_spohn}H. Spohn, J. Phys. A: Math. Gen. 16 4275 (1983) 

\bibitem {SSEP_large_dev}B. Derrida, J. L. Lebowitz, and E. R. Speer, J. Stat.
Phys. 107 775-810 (2002) 

\bibitem {other_modes}Other choices of modes, such as simple plane waves were
also tested, and also give good results, but with slower convergence.

\bibitem {KMP_large_dev}L. Bertini, D. Gabrielli and J. Lebowitz, J. Stat.
Phys. 121 843 (2005)
\end{thebibliography}
\end{document}